\documentclass[%
reprint,amsmath,amssymb,aps,prl,superscriptaddress,longbibliography,
]{revtex4-1}

\usepackage{graphicx}% Include figure files
\usepackage{dcolumn}% Align table columns on decimal point
\usepackage{bm}% bold math
\usepackage{epstopdf}
\usepackage{float}
\usepackage{braket}
\usepackage{dsfont}
\usepackage{amsmath}
\usepackage{xcolor}
\usepackage{lipsum}
\usepackage{color,soul}

\graphicspath{{figures/}}

\begin{document}

\title{High-impedance surface acoustic wave resonators}

\author{Yadav P. Kandel}
\affiliation{Department of Physics and Astronomy, University of Rochester, Rochester, NY, 14627 USA}

\author{Suraj Thapa Magar}
\affiliation{Department of Physics and Astronomy, University of Rochester, Rochester, NY, 14627 USA}

\author{Arjun Iyer}
\affiliation{The Institute of Optics, University of Rochester, Rochester, NY, 14627 USA}

\author{William H. Renninger}
\affiliation{The Institute of Optics, University of Rochester, Rochester, NY, 14627 USA}

\author{John M. Nichol}
\email{john.nichol@rochester.edu}
\affiliation{Department of Physics and Astronomy, University of Rochester, Rochester, NY, 14627 USA}

\begin{abstract}
Because of their small size, low loss,  and compatibility with magnetic fields, and elevated temperatures, surface acoustic wave resonators hold significant potential as future quantum interconnects. Here, we design, fabricate, and characterize GHz-frequency surface acoustic wave resonators with the potential for strong capacitive coupling to nanoscale solid-state quantum systems, including semiconductor quantum dots. Strong capacitive coupling to such systems requires a large characteristic impedance, and the resonators we fabricate have impedance values above 100 $\Omega$. We achieve such high impedance values by tightly confining a Gaussian acoustic mode. At the same time, the resonators also have low loss, with quality factors of several thousand at millikelvin temperatures. These high-impedance resonators are expected to exhibit large vacuum electric-field fluctuations and have the potential for strong coupling to a variety of solid-state quantum systems.

\end{abstract}

%\date{\today}\\
\pacs{}

\maketitle

\section{Introduction}
Interconnects play a key role in classical information processors and will likely play a similarly important role in large-scale quantum information processors~\cite{bravyi2022future,awschalom2021development}. Among the different types of potential quantum interconnects, mechanical resonators stand out for their desirable properties, including low loss, small size, and ability to couple to nearly all quantum devices~\cite{Aspelmeyer2003,kurizki2015quantum,aref2016quantum,chu2020perspective,clerk2020hybrid}. 
%Indeed, mechanical resonators form the basis for not only quantum interconnects, but also quantum memories~\cite{wallucks2020quantum,chamberland2022building}, frequency converters~\cite{stannigel2010optomechanical,bochmann2013nanomechanical,andrews2014bidirectional}, and even qubits. 
Piezoelectric mechanical resonators can directly interact with other quantum devices that couple to electric fields. As a result, piezoelectric resonators have found significant use in recent experiments with solid-state qubits~\cite{oconnell2010quantum,chu2017quantum,mirhosseini2020superconducting}.

The resonant, capacitive coupling $g$ between a piezoelectric mechanical resonator and a qubit is proportional to the zero-point electric-field fluctuations of the resonator. As with other types of resonators, the strength of these fluctuations is inversely proportional to the square root of the mode volume, and $g \propto 1/\sqrt{V}$~\cite{Schuetz2015}. Because the zero-point energy of a resonator does not depend on its volume, resonators with smaller volumes have larger zero-point energy densities and thus larger electric field fluctuations. Equivalently, for piezoelectric resonators, the effective \textit{parallel} RLC circuit representing the resonator should have a high characteristic impedance $Z_c=\sqrt{L/C}$, where $L$ and $C$ are the inductance and capacitance of the circuit. The resulting coupling $g \propto \sqrt{Z_c}$. Surface acoustic waves (SAWs)~\cite{Datta1986,morgan2010surface} are mechanical modes that are naturally confined to surfaces and thus present an attractive option for tightly-confined resonant modes~\cite{aref2016quantum,delsing2019}. In part because of this promise and because they are relatively easy to fabricate and measure, recent research has focused on realizing the potential of SAW resonators for solid-state quantum information processing~\cite{gustafsson2014propagating,Schuetz2015,golter2016optomechanical,manenti2016surface,moores2018cavity,satzinger2018quantum,bienfait2019phonon,sletten2019resolving,bienfait2020quantum,noguchi2017qubit,whiteley2019spin,maity2020coherent,maity2022mechanical}. 

SAW resonators feature tight confinement in the spatial direction normal to the surface, and lateral confinement can occur through phononic bandgaps~\cite{benchabane2006evidence,shao2019phononic} and focusing electrodes~\cite{kharusi1972diffraction,wilcox1985time,wilcox1985frequency,delima2003focusing,laude2008subwavelength,vainsencher2016bi,whiteley2019spin,msall2020focusing,decrescent2022large,imany2022quantum}. Despite this progress, however, the zero-point electric-field fluctuations of current SAW resonators are still not large enough for strong capacitive coupling to nanoscale qubits. For example, achieving strong coupling between a microwave resonator and a semiconductor quantum-dot qubit generally requires a characteristic impedance larger than 100 $\Omega$~\cite{Mi2018,Samkharadze2018,Landig2018}. (We refer to such an impedance as ``high," because it exceeds 50 $\Omega$, the most common impedance for microwave electronics.) High impedances such as these can be readily achieved with superconducting microwave resonators, but SAW resonators generally feature significantly lower impedance values on the order of 1-10 $\Omega$~\cite{satzinger2018quantum,saw}. 

%generally lower than what can be achieved with superconducting microwave resonators. 
In this work, we demonstrate the creation of high-impedance SAW resonators. We fabricate GHz-frequency SAW resonators whose equivalent parallel RLC circuits have characteristic impedances exceeding 100~$\Omega$ and with quality factors of several thousand at mK temperatures. We achieve such high impedance values through multiple strategies to confine the mode and boost the zero-point electric-field fluctuations. We use lithium niobate, a strong piezoelectric material as the substrate, and we use curved, highly-reflective focusing mirrors to generate a confined Gaussian mode, as well as the quasi-constant acoustic reflection periodicity geometry for the transducer~\cite{ebata1988suppression,satzinger2018quantum} to mitigate coupling to bulk acoustic modes. These results underscore the potential of SAW resonators as key elements of hybrid quantum systems.  

\section{Resonator design and fabrication}
We optimize our SAW resonators for large characteristic impedance and low dissipation, balancing several competing effects. First, we use 128$^{\circ}$ Y-cut LiNbO$_3$ as a substrate. Because the characteristic impedance of the effective parallel RLC circuit representing the SAW resonator is proportional to the piezoelectric coupling, $K^2$~\cite{satzingerthesis}, we choose 128$^{\circ}$ Y-cut LiNbO$_3$ as a substrate for its relatively high $K^2$. Second, we engineer the resonators themselves for small mode volumes. Mirrors for SAW resonators are generally thin metal electrodes or etched grooves in the substrate. The reflectivity per electrode or groove increases with the ratio of the thickness of the electrode or groove $h$ to the SAW wavelength $\lambda$, and as the reflectivity increases, the mode volume decreases. Thus, we fabricate our resonators with relatively large $h/\lambda$.
%$0.025<h/\lambda<0.135$. 
Our electrodes are electrically floating to take advantage of regeneration reflection~\cite{Datta1986}. We also fabricate small interdigital transducers (IDTs) to minimize the mode volume. %with between 3 and 11 pairs of electrodes. 
The small size also reduces the IDT capacitance, as discussed further below. 

Although thick electrodes enable small mode volumes, they also increase the coupling between surface and bulk modes, which increases dissipation~\cite{Schuetz2015}. To mitigate this effect, we use the quasi-constant acoustic reflection periodicity (QARP) geometry~\cite{ebata1988suppression,satzinger2018quantum} for the IDT. Conventional SAW resonators feature a relatively large space between the electrodes of the IDT and those of the mirrors. The QARP geometry eliminates this gap such that the electrodes have an approximately constant spacing over the length of the resonator to ensure a nearly constant effective wave speed for SAWs along the direction of propagation. Another drawback of increased electrode thickness is the significant mass loading of the resonator, which can reduce effective SAW velocity and, thus, the frequency of the resonator. To compensate for this effect, we adjust the periodicity of the electrodes to ensure that the desired resonance is in the center of the mirror stop band~\cite{decrescent2022large,suppMat}. 

\begin{figure}
    \centering
	\includegraphics[width=0.49\textwidth]{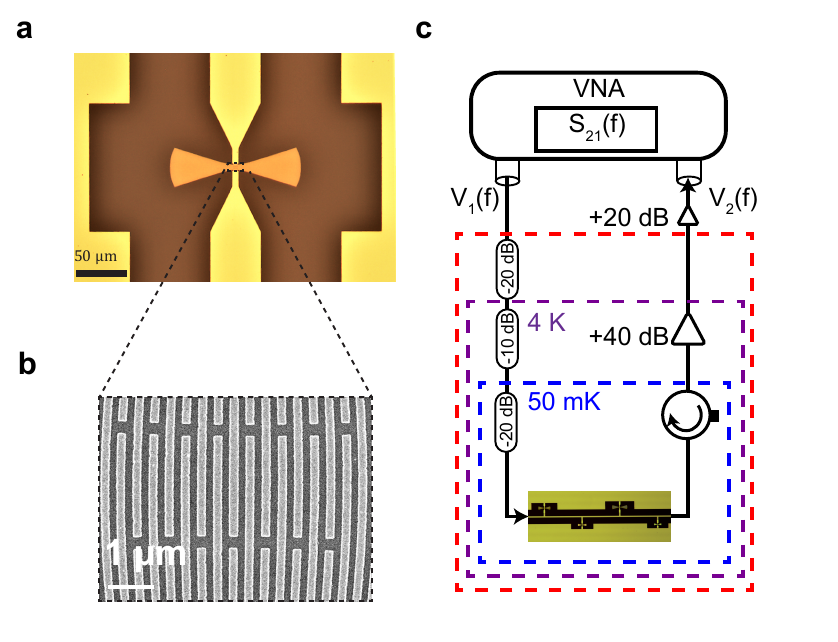}
	\caption{Gaussian SAW resonator and measurement setup.
		\textbf{a} Optical micrograph of a Gaussian SAW resonator with contact pads for probe tips. 
		\textbf{b} Scanning electron micrograph of the transducer of a typical Gaussian SAW resonator.  
		\textbf{c} Cryogenic measurement setup. Using a vector network analyzer (VNA), we measure the transmission as a function of frequency $S_{21}(f)$ through a Nb coplanar waveguide with multiple SAW resonators attached in ``hanger" mode~\cite{probst2015efficient}.}\label{fig:SAW}
\end{figure}

A natural approach to confine the mode in the transverse direction is to reduce the length of the electrodes. For flat mirrors, however, reducing their length increases diffraction losses~\cite{aref2016quantum}. To mitigate this effect, we design our mirrors to focus SAWs~\cite{de2003focusing,kharusi1972diffraction,wilcox1985frequency,wilcox1985time} and create a Gaussian mode~\cite{msall2020focusing,whiteley2019spin,ororke2020slowness} with a relatively small beam waist. (Empirically, we find that a beam waist of $2 \lambda$ enables small mode volumes with relatively large quality factors. Larger beam waists increase the mode volume, and smaller beam waists require mirrors with larger curvature, which creates additional challenges due to the anisotropy of the substrate. Here, the wavelength $\lambda \approx 1~\mu$m at 4 GHz.) We design our mirrors taking into account the simulated angle-dependent group velocity of 128$^{\circ}$ Y-cut LiNbO$_3$~\cite{suppMat}.To avoid exciting higher-order Gaussian modes, we apodize the transducer electrodes such that their length is 0.8 times the beam waist of the Gaussian mode. We empirically find that this geometry results in minimal coupling to higher-order modes. We fabricate the resonators using electron beam lithography, ultra-high-vacuum electron beam evaporation of Al, and liftoff. Figures~\ref{fig:SAW}(a)-(b) show images of a typical Gaussian SAW resonator.

\section{room-temperature characterization}
\begin{figure*}
    \centering
	\includegraphics[width=0.95\textwidth]{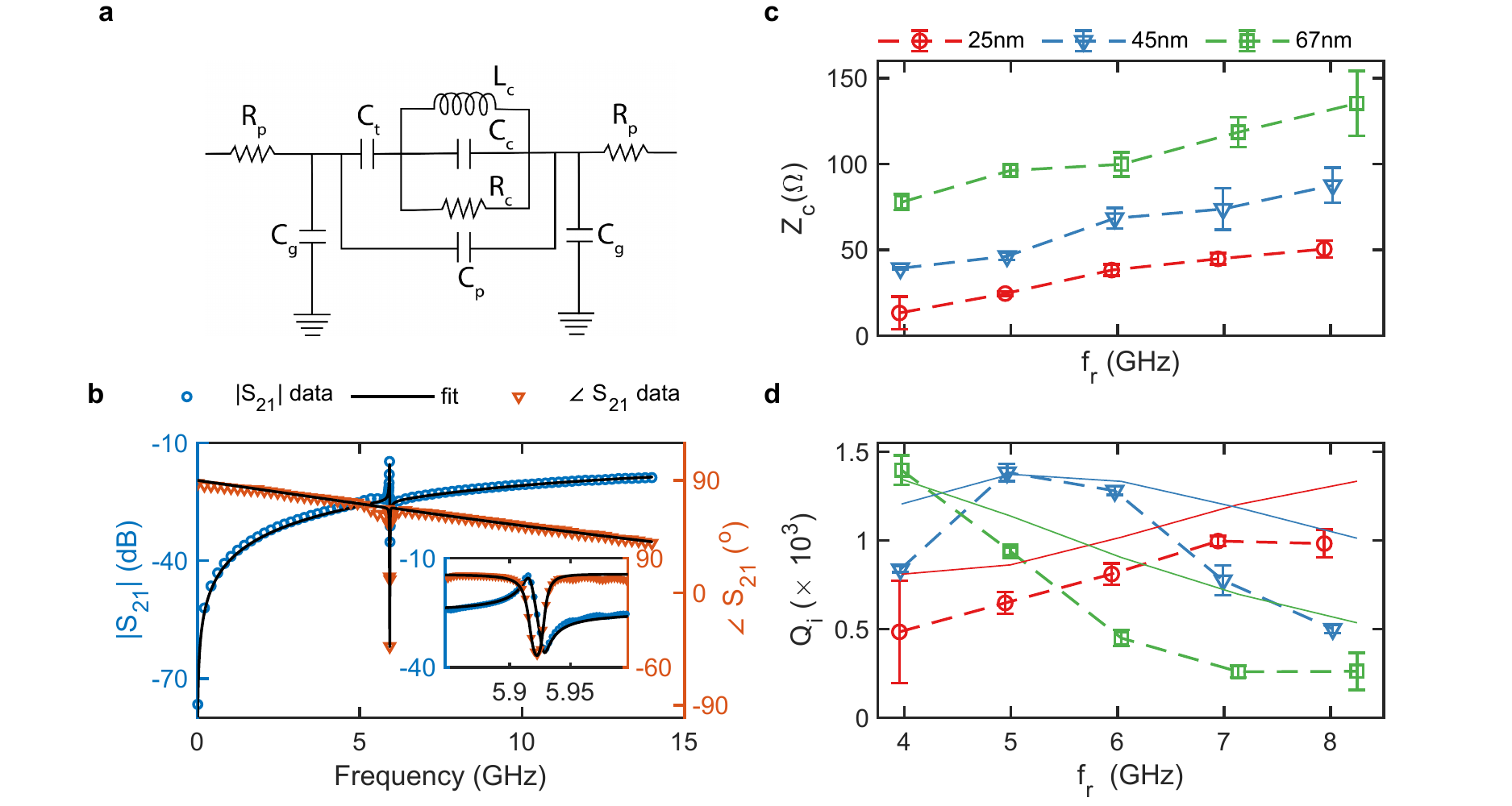}
	\caption{Room temperature characterization of SAW resonators.
		\textbf{a} The effective circuit used to model the response of the SAW resonators.
		\textbf{b} Measured magnitude and phase of the transmission through the transducer of a 6-GHz resonator and fit.  
		\textbf{c-d} Extracted characteristic impedance and quality factor values for resonators with frequencies between 4 and 8 GHz. For each frequency, 5 different resonators are characterized, and their average parameters are indicated.  The error bars are the range of the fit values. The resonators have Al electrodes of different thicknesses. Panel  \textbf{d} displays theoretical predictions for the quality factors in solid lines.
        }\label{fig:RT}
\end{figure*}

Surface acoustic wave resonators function well at room temperature. We take advantage of this property to perform a thorough electrical characterization of the SAW resonators at room temperature. We use a custom probe station with microwave probe tips and a vector network analyzer to measure the transmission through the IDT.  To facilitate this measurement, we fabricate SAW resonators with large contact pads for both the IDT and ground. Figure~\ref{fig:RT}(b) shows a typical resonance observed at room temperature. 

%The resonators have Al electrodes of different thickness.

Surface acoustic wave resonators are commonly modeled using the Butterworth-Van Dyke (BVD) circuit model, which consists of a series RLC circuit in parallel with the IDT capacitance~\cite{morgan2010surface}. Such a circuit naturally models the electrical admittance associated with SAW propagation. The BVD model is not the only way to model piezoelectric resonators. In fact, there exist mappings between the BVD model and a variety of other effective circuits, including a parallel RLC model~\cite{wollack2021loss,satzingerthesis}. The parallel RLC model consists of a parallel RLC circuit in series with the IDT capacitance [Fig.~\ref{fig:RT}(a)]. This model emphasizes the impedance of the resonator instead of its admittance. In the quantum regime, this circuit can be interpreted as a quantum harmonic oscillator in series with a coupling capacitance from the IDT. The main advantage of this model over the traditional BVD model is that the displacement of this oscillator can be identified with the voltage across the IDT pads. In the high-quality-factor limit, the zero-point voltage fluctuations are proportional to the resonator's ``characteristic impedance" $Z_c=\sqrt{L_c/C_c}$~\cite{vool2017introduction}.  

When modeling the transmission through these devices, we also include stray capacitances between the contact pads and ground ($C_g$), as well as between the contact pads themselves ($C_p$) [Fig.~\ref{fig:RT}(a)]. We need these elements to accurately model our data because the capacitance of the IDT itself is only a few femto-Farads (fF) for our devices, and the large contact pads and probe tips can easily add comparable or larger capacitances to the circuit. We also add a contact resistance ($R_p$) associated with the probe tips~\cite{satzinger2018quantum}. Using the model of Fig.~\ref{fig:RT}(a), we calculate the transmission versus frequency as a function of the circuit parameters, and we fit the data to extract the parameters. 

Fitting our data to such a complex model poses multiple challenges, including determining the large stray capacitances, which can overwhelm the transducer capacitance. As described further in the Supplemental Material~\cite{suppMat}, we solve this challenge by fabricating a series of control devices for each resonator. One control device has the same contact pads but no IDT or mirror electrodes. The other control device has contact pads and IDT electrodes but no mirror electrodes.  Using finite element simulations of the device geometry to produce initial guesses for the stray capacitances, we fit the data from the first control device (contact pads only, no IDT or mirrors) to extract $C_p$, and we determine $C_g$ through finite element simulations.  We fit the data from the second control device (contact pads and IDT, but no mirrors) to extract the IDT capacitance, and we fit the full devices to extract the resonator parameters (Fig.~\ref{fig:RT}). We have confirmed that our fitting results do not change significantly if the initial values of $C_p$ and $C_g$ are changed by $\pm 25 \%$ from the values predicted by finite-element simulation. As discussed further below, we also corroborate our fit results with three-dimensional finite-element simulations of microwave transmission through the resonators. 

Figure~\ref{fig:RT}(b) displays the transmission through a typical 6-GHz resonator with 200 45-nm-thick Al electrodes in Gaussian mirrors on either side of the transducer and our fits. Figure~\ref{fig:RT}(c) displays our extracted impedance values for resonators with frequencies between 4 and 8 GHz with different electrode thicknesses. For each frequency, we measure five nominally identical resonators, and we plot the average values. We observe that as the frequency of the resonator increases, the effective electrical impedance increases, consistent with our expectation that the mode volume decreases with the acoustic wavelength. We also observe that for a fixed frequency, as the electrode thickness increases, the impedance increases, consistent with our expectation that thicker electrodes confine the mode more effectively. We achieve a maximum impedance of well above 100 $\Omega$ for resonators in the frequency range tested. 

Typical values of the IDT capacitance, $C_t$, in our devices are about 5-10 fF. Typical stray capacitance values are $C_p \sim 10$ fF and $C_g\sim 70$ fF~\cite{suppMat}. Future hybrid quantum devices using capacitive coupling to high-impedance SAW resonators will require low-capacitance wiring with $C_p, C_g \ll C_t$. The most significant contribution to the stray capacitance in our devices comes from the large probe-tip contact pads, which can be eliminated in future devices. Without the large contact pads, we expect that low-capacitance wiring to other solid-state systems can easily be achieved~\cite{harvey2022coherent}.

Figure~\ref{fig:RT}(d) displays the extracted internal quality factors $Q_i=R_c\sqrt{C_c/L_c}=R_c/Z_c$. For 25-nm-thick electrodes, we observe that the quality factor increases with frequency. However, for 67-nm-thick electrodes, we observe that the quality factor decreases with frequency, and for 45-nm-thick electrodes, we observe a maximum in the quality factor around 5 GHz, and then a reduction in the quality factor at higher frequencies. We can understand these trends by considering the competition between loss through the mirrors, which decreases with $h/\lambda$, and bulk dissipation, which increases with $h/\lambda$~\cite{Schuetz2015}. Figure~\ref{fig:RT}(d) also displays theoretical predictions calculated using expressions for grooved, flat resonators~\cite{Schuetz2015} for the total quality factor considering mirror loss, bulk loss, and material loss. The only free parameters in our calculations are a scaling factor (0.2) for the predicted reflectivity per electrode~\cite{Datta1986}, to account for the angle-dependent piezoelectric coupling in the substrate, and a quality factor associated with material loss~\cite{suppMat}. The relatively good agreement between our data and the predictions corroborates the interpretation of the different trends in the data.

\begin{figure}
    \centering
	\includegraphics[width=0.5\textwidth]{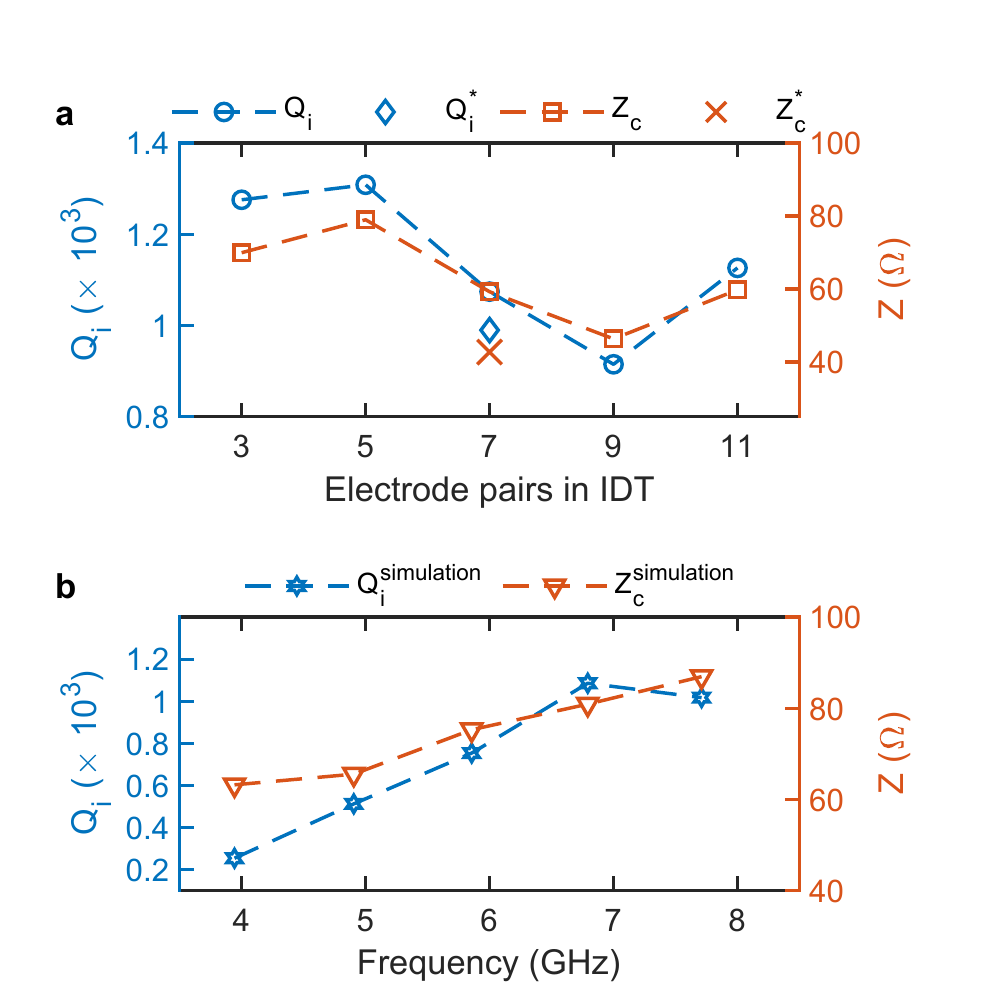}
	\caption{Comparison of different resonator types.
		\textbf{a} Comparison of 6-GHz Gaussian resonators with different numbers of electrodes in the IDT. For IDTs with 7 pairs of electrodes, data for non-QARP Gaussian resonators, indicated by an $^*$, are also shown. Electrodes are 45 nm thick.  
        \textbf{b} Results of finite element simulations of SAW resonators with 45-nm-thick Al electrodes and seven pairs of electrodes in the IDT.} \label{fig:control}
\end{figure}

We have checked that our various strategies to confine the mode have the desired effect (Fig.~\ref{fig:control}). For example, increasing the number of electrodes in the IDT for a fixed thickness should increase the mode volume and thus decrease the impedance, in agreement with our observations [Fig.~\ref{fig:control}(a)]. We have also fabricated resonators with the standard electrode periodicity as opposed to the QARP geometry. In these cases, we find reduced impedance values and quality factors, presumably due to increased coupling to bulk modes [Fig.~\ref{fig:control}(a)]. Finally, we have fabricated SAW resonators with flat mirrors and similar sizes to the Gaussian resonators. The smallest flat resonators (beam width equal to 4$\lambda$) are apparently not effective in confining a single mode, because multiple modes spaced by less than the free spectral range are visible~\cite{suppMat}. For larger flat resonators (beam width equal to 42$\lambda$), we observe single modes with quality factors significantly lower than the Gaussian resonators, likely because of increased diffraction loss~\cite{suppMat}. Note that a beam width of 42$\lambda$ is significantly smaller than other flat SAW resonators reported in the literature, which are typically on the order of hundreds of wavelengths wide~\cite{satzinger2018quantum}.

We have also conducted three-dimensional finite element simulations in COMSOL Multiphysics [Fig.~\ref{fig:control}(b)]. As described further in the Supplemental Material~\cite{suppMat}, we simulate the microwave transmission through the transducer for SAW devices with different resonance frequencies.  For each SAW resonator, we fit the simulated transmission to the circuit of Fig.~\ref{fig:RT}(a), except that we do not include any stray capacitances or contact resistance, because the simulation contains no probe tips or contact pads. Figure~\ref{fig:control}(b) shows the quality factors and impedance values from these fits. The simulated impedance increases with the frequency, in quantitative agreement with our measurements. The quality factor reaches a maximum near 7 GHz. This trend agrees qualitatively with our data, which indicates that the quality factor goes through a maximum in this frequency range. While the measured devices have 200 electrodes in each mirror, the simulated devices have only 50 electrodes to reduce the computational resources required. We hypothesize that energy loss through the mirrors reduces the quality factor for low-frequency devices more significantly in our simulations than in the actual devices. It could also be that the specific details of the finite element simulation, such as the mesh density, which is constrained by our computational resources, contribute to an inaccurate determination of the quality factor.  

These measurements and simulations highlight the potential of Gaussian SAW resonators for small mode volumes and correspondingly high impedance values. Based on our finite-element simulations, we expect that our mode volumes are on the order of a few cubic $\mu$m. For our highest-frequency resonators, the mode volume is significantly smaller than what can be achieved with phononic-crystal mirrors, for example~\cite{shao2019phononic}. Moreover, our approach also relies on a simple, intuitive analogy with free-space optics to confine the mode. The ability to create resonators with effective electrical impedances well above 50 $\Omega$ opens the door to strong coupling to various quantum systems. 

\section{Cryogenic characterization}

\begin{figure*}
    \centering
	\includegraphics[width=0.95\textwidth]{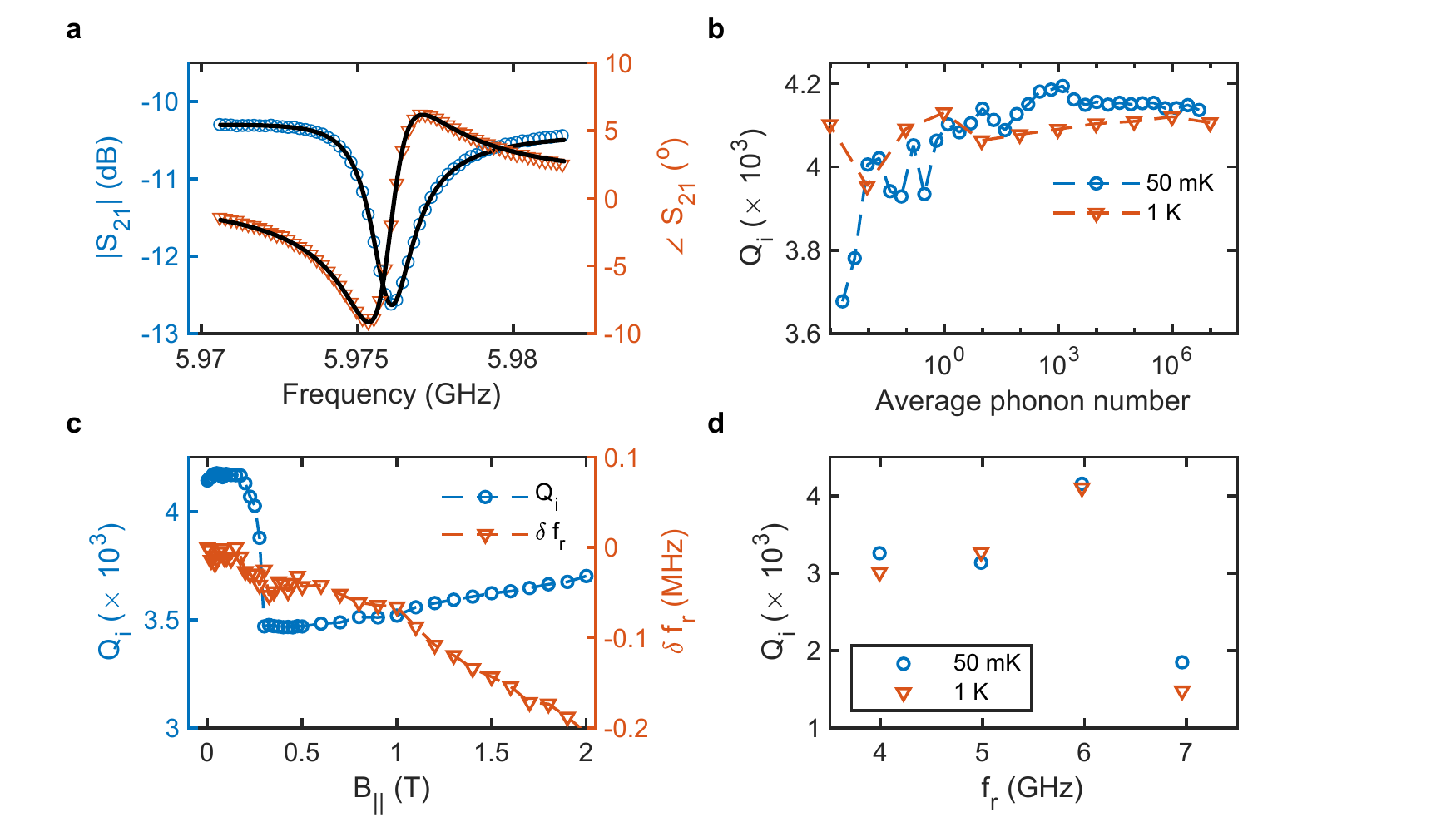}
	\caption{Millikelvin characterization of SAW resonators. 
		\textbf{a} Magnitude and phase of the transmission through the coplanar waveguide near 6 GHz and fit for the 6-GHz resonator.
		\textbf{b} Internal quality factors vs. average phonon number for the 6-GHz resonator.
		\textbf{c} Internal quality factor and change in resonance frequency compared to zero magnetic field value $\left(\delta f_r = f_r\left(B_{||}\right) - f_r(0)\right)$ versus in-plane magnetic field for the 6-GHz resonator. 
        \textbf{d} Internal quality factor vs resonator frequency. Data for all panels (a), (c), and (d) were taken in the large phonon-number limit. }\label{fig:LT}
\end{figure*}

Having confirmed the potential for small mode volumes at high frequencies at room temperature, we characterize the resonators at mK temperatures in a dilution refrigerator. We fabricated a niobium coplanar waveguide transmission line on the LiNbO$_3$ substrate, and then we coupled multiple resonators of different frequencies to the waveguide. One side of each IDT is galvanically connected to the center conductor, and the other side is galvanically connected to the ground plane. In this configuration, we measure the resonators in ``hanger" mode~\cite{probst2015efficient} [Fig.~\ref{fig:SAW}(c)]. 

Figure~\ref{fig:LT}(a) shows a typical resonance curve and fit associated with a 6-GHz resonator with 45-nm-thick Al electrodes. Because extracting the effective circuit parameters (as we did at room temperature) relies critically on accurate calibration of the entire microwave circuit, and because of the challenges associated with calibrating the microwave wiring in a cryostat, we cannot directly determine the impedance of the resonators at low temperature. Instead, we focus on measuring the resonator quality factor at low temperature ~\cite{probst2015efficient}. Previous studies have suggested that some piezoelectric coupling coefficients in LiNbO$_3$ do not depend very strongly on temperature~\cite{chen2021investigation,islam2019piezoelectric}, so we hypothesize that the impedance does not change significantly on cooling to cryogenic temperatures. 
We observe only slight (approximately 90 MHz) frequency shifts when cooling down the resonators. These shifts may result from thermal contraction or a  decrease in the dielectric constant, which would affect the transducer capacitance. Nevertheless, the minimal frequency shifts we observe suggest that the effective inductance and capacitance of the resonator, and thus its impedance, do not change significantly upon cooling. Future studies can also confirm this by measuring the frequency difference between the admittance and impedance maxima of the resonator at different temperatures~\cite{satzingerthesis}.

We measure the transmission through the coplanar waveguide and fit the data near each resonance to a standard resonator model, which features the resonance frequency, and internal and coupling quality factors ($Q_i$ and $Q_c$)as fit parameters~\cite{probst2015efficient}. Figure~\ref{fig:LT}(b) shows the extracted internal and coupling quality factors for the 6-GHz resonator at different average phonon numbers calculated as $\langle n\rangle = \frac{Q_c}{\omega_r} \left( \frac{Q_i}{Q_i + Q_c}\right)^2 \frac{P_{in}}{\hbar \omega_r}$~\cite{yu2021magnetic} where $\hbar$ is the reduced Planck constant, $\omega_r = 2\pi f_r$, and $P_{in}$ is the input power. We observe that the quality factor increases weakly with phonon number at both 50 mK and 1 K. This behavior suggests the saturation of two-level systems~\cite{manenti2016surface}, as is often seen with superconducting microwave resonators.  We also plot the internal quality factor and change in resonance frequency versus the in-plane magnetic field [Fig.~\ref{fig:LT}(c)]. We observe that the internal quality factor drops by a small amount near a few hundred mT. This behavior is consistent across all resonators we measure, and we hypothesize that this reduction occurs when the Al electrodes become non-superconducting. We expect a much higher critical field for the Nb coplanar waveguide. (We also observed slight [$\sim 5$ MHz] irreversible downward shifts in the resonator frequencies on the first field sweep. See the Supplemental Material for more details~\cite{suppMat}.)  Despite this small drop, the resonator, which is expected to have an impedance around 100 $\Omega$ based on room-temperature measurements, maintains a quality factor above 3000 until a magnetic field of 2 T, an encouraging prospect for compatibility with devices such as spin qubits, which often require magnetic fields for their operation. 

Finally, we plot the internal quality factor for different resonators at both 50 mK and 1 K at zero magnetic field [Fig.~\ref{fig:LT}(d)]. Resonators with frequencies between 4 and 6 GHz have internal quality factors of several thousand at mK temperatures. As above, with the room temperature resonators, we observe that the quality factor decreases for the highest-frequency resonators. In contrast to the room-temperature devices, the 6 GHz resonator maintains a relatively high quality factor, while the 8-GHz resonator has a quality factor that was too low to measure.  These differences may result from variations in fabrication conditions between devices. (The cryogenic devices were made during a different fabrication run and with a different process than the room temperature devices~\cite{suppMat}.) 

\section{Outlook}
We have demonstrated that through a combination of design choices, we can create SAW resonators with high impedance values and quality factors at mK temperatures. The performance of the SAW resonators described here is already sufficient to enable strong coupling between an individual electron in a quantum dot~\cite{mi2017strong,Stockklauser2017,scarlino2022situ} and a single phonon~\cite{suppMat}. To estimate the charge-phonon coupling rate possible with our current devices, consider the results of Ref.~\cite{scarlino2022situ}, where a charge-photon coupling rate $g_c/(2 \pi)$=154 MHz was measured between a 5-GHz, 1-k$\Omega$ 
superconducting resonator and a single electron in a double dot with a corresponding charge-qubit dephasing rate $\gamma_c/(2\pi)=28.3$ MHz. Because the charge coupling rate scales as the square root of the impedance, and taking a typical SAW-resonator impedance of 100 $\Omega$, we would expect a charge-phonon coupling of $g_{ph}/(2 \pi)=\sqrt{\frac{1}{10}}g_{c}/(2 \pi)$= 47 MHz, which is still larger than the charge-qubit dephasing rate.  Taking a typical SAW-resonator quality factor of 3000, the phonon decay rate is expected to be $\kappa_{ph}/(2 \pi) = f/Q \sim 2$ MHz, which is less than $g_{ph}/(2\pi)$. In order to achieve strong spin-phonon coupling, even larger impedances will be required. Alternatively, a high-impedance superconducting microwave resonator could be used as an intermediary between a mechanical resonator and spin qubit to boost the coupling, similar to the case where two qubits coupled to the same resonator experience an effective coupling~\cite{majer2007coupling}. 
 
A major technical challenge to the integration of mechanical resonators and quantum-dot devices is to suppress excess capacitance from the resonator-qubit wiring. Reducing stray capacitance is also a challenge in coupling quantum dots to superconducting resonators. Encouragingly, however, low-capacitance resonators have already been successfully integrated with quantum dots~\cite{harvey2022coherent,scarlino2022situ}. Another technical challenge involves the integration of dissimilar materials, such as LiNbO$_3$ with spin-qubit materials like Si/SiGe or GaAs. We envision overcoming this challenge with flip-chip integration techniques, which have been successfully demonstrated with various qubit platforms~\cite{satzinger2019simple,holman20213d}. Lithium niobate thin films can also be bonded to other materials like silicon~\cite{luschmann2023surface}. 

Beyond semiconductor qubits, we expect that the high-impedance SAW resonators presented here will also find use in other platforms. Capacitive coupling between superconducting qubits and SAW resonators have enabled exploring dissipation-mediated state preparation~\cite{kitzman2023phononic}, mode-selective coupling~\cite{moores2018cavity}, strong-dispersive couplings~\cite{sletten2019resolving}, and quantum cavity acoustodynamics~\cite{manenti2016surface}. We envision that high-impedance resonators can enable further research along these directions. High-impedance SAW resonators may also enable coupling to other types of voltage-tunable spin qubits, like self-assembled quantum dots~\cite{tsuchimoto2017proposal,tsuchimoto2022low}. %Direct spin-phonon couplings have been demonstrated in such systems, but capacitive coupling may eliminate the need for placing the qubit itself in the cavity. 
For all of these applications, a detailed assessment of the dissipation mechanisms will be helpful for further progress toward increasing the impedance and reducing the dissipation. Our measurements show a relatively small reduction in quality factor when the Al electrodes become non-superconducting, suggesting that electrical resistivity in the transducer is not the dominant loss mechanism at low temperatures. Given the reduction in quality factor with frequency observed in Fig.~\ref{fig:RT}, we hypothesize that coupling to bulk modes is a limiting mechanism, potentially resulting from the details of the electrode curvature. In addition, the metal electrodes in the transducer also likely have a significant impact on the reflections in the cavity. Exploring, for example, strategies to minimize reflection associated with the transducer, such as using a split-finger geometry, could also be worthwhile. 

Altogether, our results emphasize the potential of SAW resonators for integration in future hybrid architectures. Just as with superconducting microwave resonators, whose compatibility with a variety of quantum devices is enhanced by the potential of high impedance and large vacuum electric field fluctuations, the ability to engineer confined acoustic modes with low dissipation opens the door to exciting applications for hybrid quantum architectures.   

\section{Data Availability}
The processed data are available at \href{https://doi.org/10.5281/zenodo.8060648}{https://doi.org/10.5281/zenodo.8060648}~\cite{yadav_p_kandel_2023_8060648}. The raw data are available from the corresponding author upon reasonable request.

%\bibliography{SAWbib}
%\nocite{mcrae2020materials,Kouwenhoven2001}
\section{Acknowledgments}
This work was sponsored by the Army Research Office, grant number W911NF-19-1-0167, and the Office of Naval Research, grant number N00014-20-1-2424. The views and conclusions contained in this document are those of the authors and should not be interpreted as representing the official policies, either expressed or implied, of the Army Research Office, the Office of Naval Research, or the U.S. Government. The U.S. Government is authorized to reproduce and distribute reprints for government purposes notwithstanding any copyright notation herein.

\end{document}

\title{Supplemental Material for \\ ``High-impedance surface acoustic wave resonators"}

\author{Yadav P. Kandel}
\affiliation{Department of Physics and Astronomy, University of Rochester, Rochester, NY, 14627 USA}

\author{Suraj Thapa Magar}
\affiliation{Department of Physics and Astronomy, University of Rochester, Rochester, NY, 14627 USA}

\author{Arjun Iyer}
\affiliation{The Institute of Optics, University of Rochester, Rochester, NY, 14627 USA}

\author{William H. Renninger}
\affiliation{The Institute of Optics, University of Rochester, Rochester, NY, 14627 USA}

\author{John M. Nichol}
\affiliation{Department of Physics and Astronomy, University of Rochester, Rochester, NY, 14627 USA}
\email{john.nichol@rochester.edu}

%\date{\today}\\

\maketitle
	
%\date{\today}

\section{Resonator design}
To illustrate the design of our SAW resonators, consider a Gaussian beam propagating in the $x$ direction in the $x-y$ plane with the focus at $(x,y)=(0,0)$. In an isotropic medium, the radius of curvature of the wavefronts of the beam as a function of $x$ is $R(x)=x(1+(x_R/x)^2)$. The Rayleigh range $x_R=\pi w_0^2/\lambda$, where $w_0$ is the minimum beam waist, and $\lambda$ is the wavelength. For our devices, $w_0=2 \lambda$. For such an isotropic medium, the electrode edges should be chosen to follow the wavefronts. In an anisotropic medium, the electrode edges are generally chosen to follow group-velocity contours~\cite{kharusi1972diffraction,wilcox1985time}. To generate the group velocity as a function of propagation direction $\theta$ away from the $x$ direction on 128$^{\circ}$ Y-cut LiNbO$_3$, we use the simulated phase velocity from Ref.~\cite{ororke2020slowness} and follow Ref.~\cite{wilcox1985time} to compute the angle-dependent group velocity $v_g(\theta)$. For an electrode crossing the $x$-axis at position $x$, the coordinates of that electrode away from the $x$-axis are $x'=x-R(x)\left(1-\frac{v_g(\theta)}{v_g(0)}\cos(\theta)\right)$ and $y'=R(x)\frac{v_g(\theta)}{v_g(0)}\sin(\theta)$. The positions of the reflectors crossing the $x$ axis are determined from the position-dependent phase of a Gaussian beam $\phi(x)=\frac{2 \pi}{\lambda}x -\frac{1}{2}\arctan \left( \frac{x}{x_R}\right)$. Away from the $x$ axis, we extend the electrodes to a distance equal to twice the beam waist $w(x)=w_0 \sqrt{1+\left(\frac{x}{x_R} \right)^2}$.

\begin{figure}
    \centering
    \includegraphics{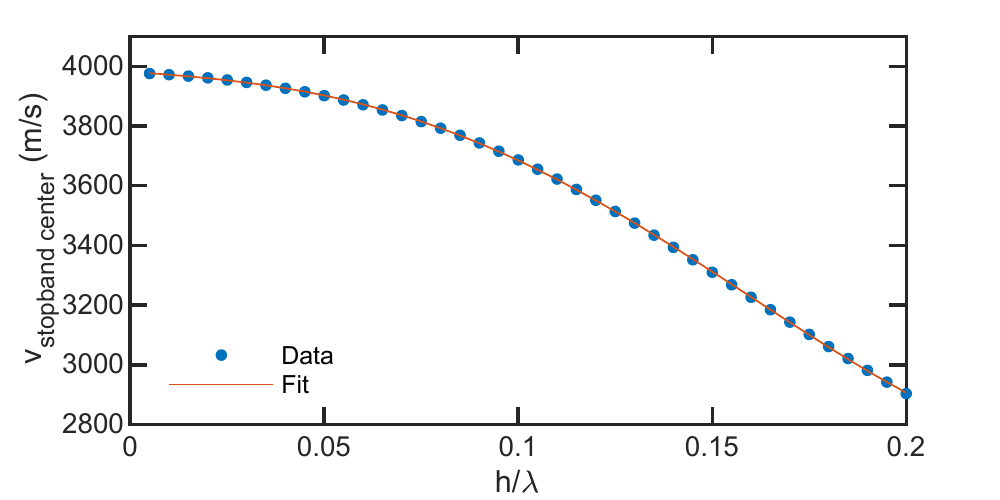}
    \caption{Simulated velocity at the center of the SAW mirror stopband as a function of the ratio of the electrode thickness to the wavelength $(h/\lambda)$. In the simulations, the ratio of the electrode width to the gap between neighboring electrodes (metalization ratio) is 1. }
    \label{fig:stopband_center}
\end{figure}

We choose the quasi-constant acoustic reflection periodicity to minimize coupling to bulk modes. The transducer electrodes extend to $0.8 \times w(x)$ to minimize the excitation of higher-order Gaussian modes. Finally, we adjust the periodicity of the metal electrodes to compensate for mass-loading that occurs with relatively thick electrodes. Following Ref.~\cite{decrescent2022large}, we simulate the frequency splitting between symmetric and antisymmetric SAW modes for a given ratio of the electrode thickness to the wavelength $h/\lambda$. These frequencies give the boundaries of the mirror stop band. As $h/\lambda$ increases, the stopband center frequency drops. In our devices, we adjust the electrode periodicity to counteract this drop to have the actual resonance frequency match the target value. Simulation data and its fit to a fourth-order polynomial is shown in Fig.~\ref{fig:stopband_center}.
% where $10^{-5} v(t/\lambda) = 9.29 (t/\lambda)^4 - 3.02 (t/\lambda)^3 + 0.02 (t/\lambda)^2 - 0.01 (t/\lambda) + 0.04 (m/s)$.

\section{Fabrication}
We fabricate our SAW resonators on single-side-polished black 128$^{\circ}$ Y-cut LiNbO$_3$ substrates using a liftoff process. We spin PMMA and DisCharge H$2$O on the substrate. We pattern the electrodes using electron-beam lithography. The exposed resist is then developed using a 3:1 mixture of isopropanol and methyl isobutyl ketone at 2$^\circ$ C. After that, we deposit Al in an ultra-high vacuum electron-beam evaporator system with a starting base pressure in the mid $10^{-10}$ Torr range. After deposition, we sonicate the substrate in a warm acetone bath to remove excess resist and metal. 

For room-temperature measurements, we create 200-nm thick Al probe-tip contact pads using a liftoff process with photolithography. We find that thick pads are required to minimize the contact resistance between our probe tips and the devices.  For cryogenic measurements, we fabricate Nb coplanar waveguides on the substrates using a liftoff process with photolithography and electron beam evaporation prior to SAW fabrication. After each metal film deposition step, we clean the substrates in an oxygen plasma to remove resist residue. 
    
\section{Fitting}
\begin{figure}
    \centering
    \includegraphics[width=0.25\textwidth]{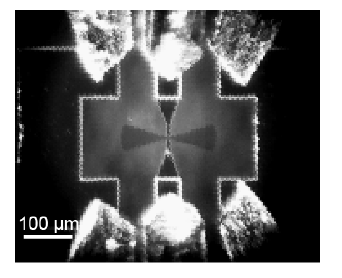}
    \caption{Optical microscope image of a SAW resonator being measured at the calibrated probe station.}\label{fig:RTmeasurement}
\end{figure}

To extract the circuit parameters of a SAW resonator, we measure it using a probe station with microwave probe tips and a vector network analyzer (VNA). Prior to measurement, we calibrate the VNA and attached cables using the short-load-open-through procedure with a calibration substrate. Figure~\ref{fig:RTmeasurement} shows a two-port SAW resonator being measured, where the probe tips connected to the VNA are in electrical contact with the resonator. 

\begin{figure}
    \centering
    \includegraphics[width=1.0\textwidth]{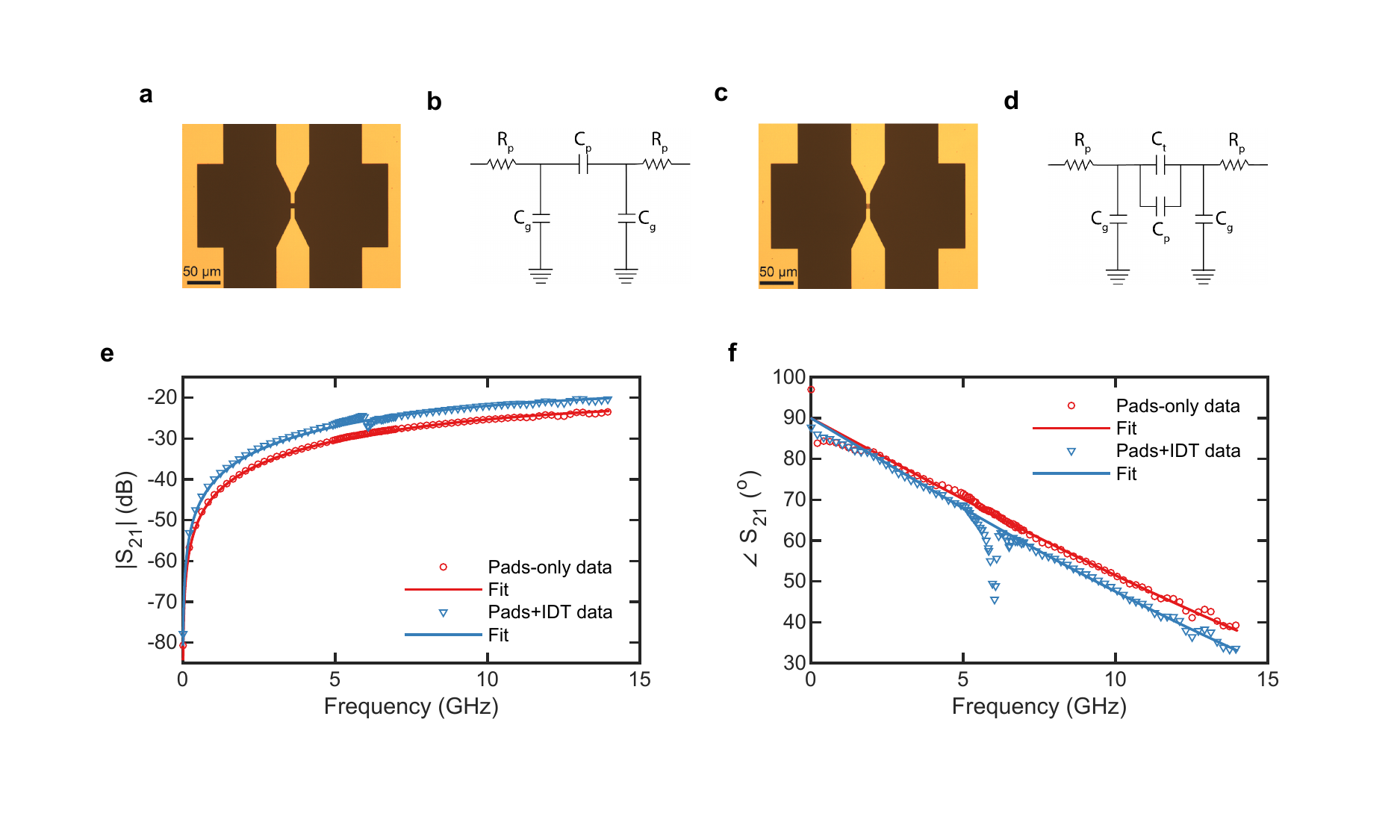}
    \caption{
        (a) Optical microscope image of a pads-only test device.  
        (b) Electrical circuit model for the device shown in (a). 
        (c) Optical microscope image of a test device with pads and an IDT. 
        (d) Electrical circuit model for the device shown in (c). Compared to (b), a capacitor is added to account for the capacitance of the IDT. 
        Measured transmission (e) magnitude and (f) phase,  and their fits for test devices. For clarity, not all data points are shown here. For pads-only device data fits (red), $R_p$ and $C_p$ are the fit parameters while $C_g$ is obtained from a finite element simulation of the exact geometry. For this particular fit, $R_p = 22.4\ \Omega$, and $C_p = 9.6$ fF.  Although piezoelectric transduction near 6 GHz is observed in data from the device with pads and IDT, the fits to the pads-and-IDT test device only account for the electrical capacitance. For the $C_p$ and $C_g$ values obtained from the pads-only device fits (red), we extract $C_t=4.9$ fF from fits to the pads-and-IDT test device (blue). See Fig.~\ref{fig:fitParamsForFig2cd} for a complete set of fit parameters.\label{fig:PadOnly_PadAndIDT} }
\end{figure}

To make extracting the different circuit parameters easier, we fabricate and measure two test devices for each different resonator. The first device has only the contact pads [Fig.~\ref{fig:PadOnly_PadAndIDT}(a)]. The second test device has the contact pads and the transducer [Fig.~\ref{fig:PadOnly_PadAndIDT}(c)]. We model the pads-only test device with the circuit of Fig.~\ref{fig:PadOnly_PadAndIDT}(b), where $R_p$ accounts for contact and film resistance, $C_p$ represents the capacitance between two transducer pads, and $C_g$ stands for the capacitance between the transducer pads and ground. In the absence of piezoelectric coupling, the transducer forms  a capacitor. We model the second test device, which includes the transducer electrodes, using the circuit of Fig.~\ref{fig:PadOnly_PadAndIDT}(d), where $C_t$ represents the capacitance of the IDT. For both test devices and the full resonator itself, we measure the transmission in both a wide frequency range (100 MHz - 14 GHz) and also a narrow frequency range (1 GHz) near the resonance. Finally, we fit the data to extract the circuit parameters in the following way:\\

\noindent \textbf{Step-0}: We simulate the pads-only test device to obtain the estimated values of circuit parameters $C_p$ and $C_g$ in Fig.~\ref{fig:PadOnly_PadAndIDT}(b). \\

\noindent \textbf{Step-1}: We fix $C_g$ to the values obtained from Step-0 and fit the wide-frequency  data from the pads-only test device to the circuit model in Fig.~\ref{fig:PadOnly_PadAndIDT}(b) with $C_p$ and $R_p$ as the fit parameters.\\

\noindent \textbf{Step-2}: We fix $C_g$ and $C_p$ to the values from the previous step and fit the wide-frequency data from pads-and-IDT test device to the circuit model in Fig.~\ref{fig:PadOnly_PadAndIDT}(d) with $C_t$ and $R_p$ as the fit parameters.\\

\noindent \textbf{Step-3}: We update $C_t$ and $R_p$ by fitting the wide-frequency data from the resonator to the circuit model in Fig.~\ref{fig:PadOnly_PadAndIDT}(d).\\

\noindent \textbf{Step-4}: We fix $C_g$, $C_p$, and $R_p$ to the values from the previous step and update $C_t$ by fitting data from the resonator around its resonance frequency to the circuit model in Fig.\ref{fig:PadOnly_PadAndIDT}(d) which ignores the resonance feature itself and fits only the background. \\

\noindent \textbf{Step-5}: We fix $C_g$, $C_p$, and $R_p$ to the values from the previous step and fit data from the resonator near its resonance frequency to the circuit model in Fig.~2(a) in the main text for $C_t,\ f_r,\ R_c,\text{and}\ C_c$.\\

\noindent \textbf{Step-6}: We update $C_t,\ f_r,\ R_c,\ \text{and}\ C_c$ by fitting data from the resonator near its resonance frequency to the circuit model in Fig.~2(a) in the main text with values from step-5 as the initial guess.\\

\begin{figure}
    \centering
    \includegraphics[width=1.0\textwidth]{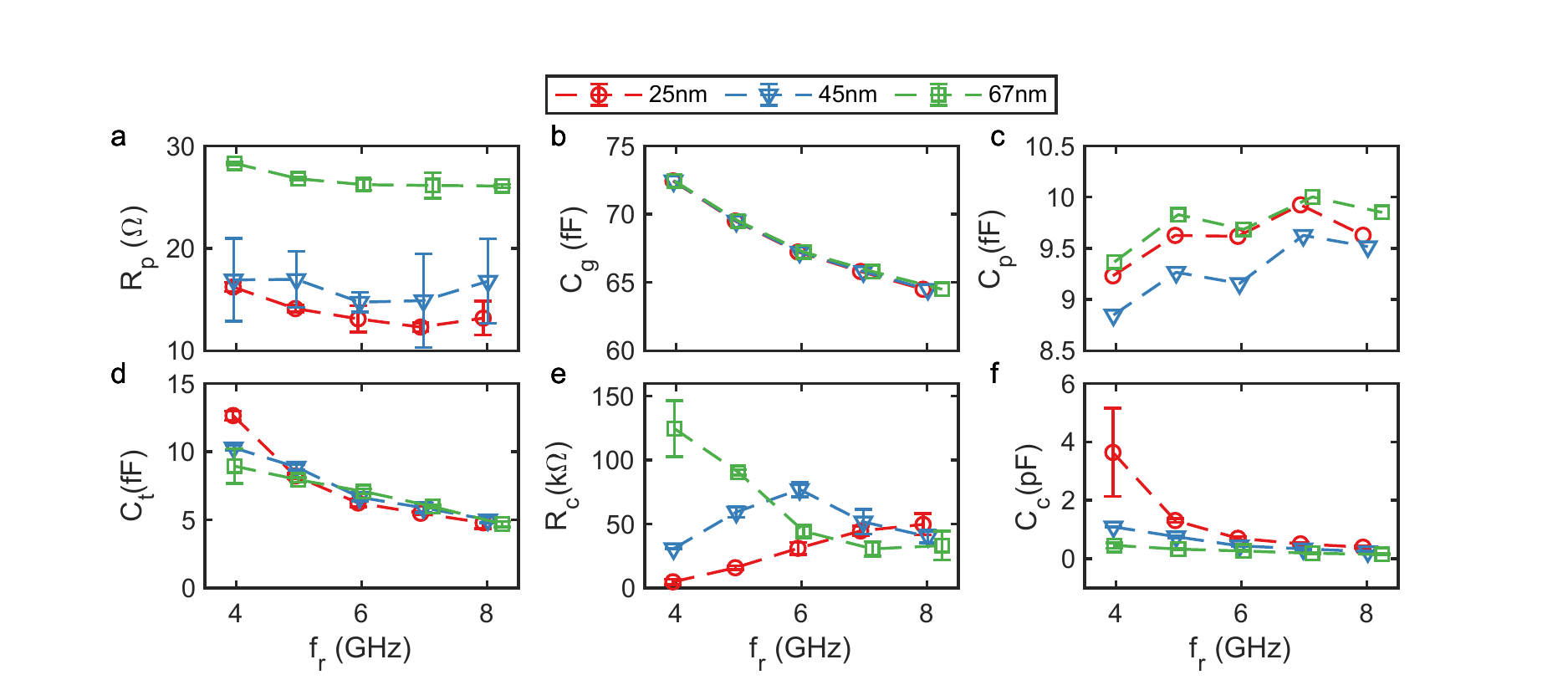}
    \caption{Fit parameters for SAW resonators of different film thicknesses, corresponding to Fig.(2) in the main text. 
        (a) Contact resistance $R_p$ vs frequency.  
        (b) Ground capacitance $C_g$ vs frequency.  
        (c) Shunt capacitance $C_p$ vs frequency.  
        (d) Transducer capacitance $C_t$ vs frequency. 
        (e) Resonator resistance $R_c$ vs frequency. 
        (f) Resonator capacitance $C_c$ vs frequency. 
        The data points in (a), (d)-(f) are the average of results from fits of 5 identical devices, and the error bars represent the range of the fit results. All parameters are plotted against the average value of the fit frequency.}
 \label{fig:fitParamsForFig2cd}
\end{figure}

The characteristic impedance $Z_c=\frac{1}{2\pi f_0 C_c}$ and the quality factor $Q_i = \frac{R_c}{Z_c}$ of the resonators are calculated from the fit values of the circuit parameters. Circuit fit parameters corresponding to $Z_c$ and $Q_i$ in Fig.~2(c-d) are shown in Fig.~\ref{fig:fitParamsForFig2cd}, where data points are the mean of the results and error bars represent the range of the parameters.

% \begin{figure}
%     \centering
%     \includegraphics[width=1.0\textwidth]{figures/Fig2_sup2_m.pdf}
%     \caption{
%         (a) Optical microscope image of a test device with pads and an IDT. 
%         (b) Electrical circuit model for the device shown in (a). Compared to Fig.~\ref{fig:PadFit}(b), a capacitor is added to account for the capacitance of the IDT. 
%         (c) Measured transmission and fits for a similar test device. For clarity, not all data points are shown here. Although piezoelectric transduction near 6 GHz is observed, this fit only accounts for the electrical capacitance. For the $C_p$, and $C_g$ obtained in~\ref{fig:PadFit}(c), $C_t$ for this specific fit example is 4.9 fF.}
%  \label{fig:IDTFit}
% \end{figure}

\begin{figure}
    \centering
    \includegraphics[width=1.0\textwidth]{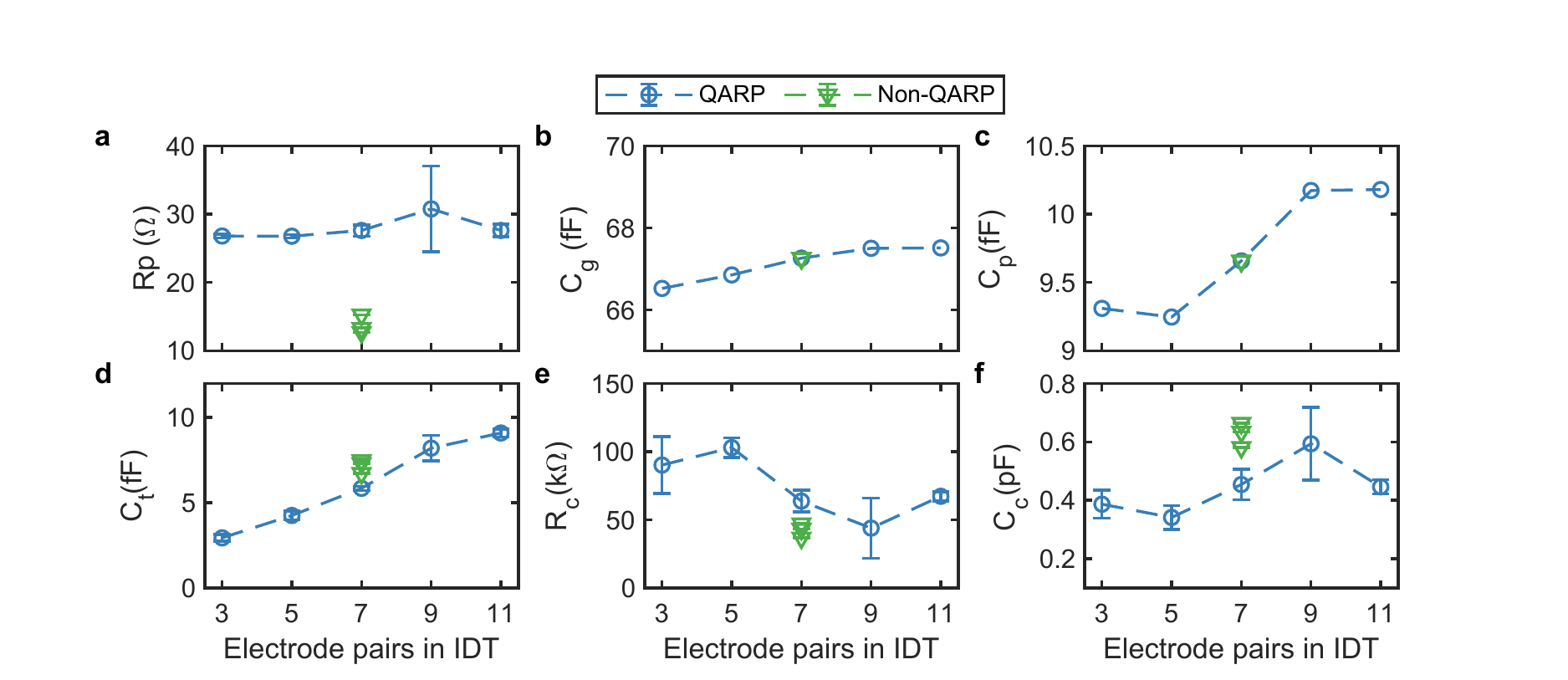}
    \caption{Fit parameters for SAW resonators with Gaussian mirrors and different numbers of electrode pairs in their IDTs, corresponding to Fig.3(a) in the main text. 
        (a) Contact resistance $R_p$ vs frequency.  
        (b) Ground capacitance $C_g$ vs frequency.  
        (c) Shunt capacitance $C_p$ vs frequency.  
        (d) Transducer capacitance $C_t$ vs frequency. 
        (e) Resonator resistance $R_c$ vs frequency. 
        (f) Resonator capacitance $C_c$ vs frequency.         
    The data points in (a), (d)-(f) are the average of fit results from 3 identical devices, and the error bars represent the spread of the fit results. In all figures, circles are for the devices with QARP design, and wedges are for the devices without QARP design.}
    \label{fig:pBVDparamsForQARP-nonQARP}
\end{figure}

\section{Quality factor prediction}
Following Ref.~\cite{Schuetz2015}, we calculate loss due to leakage out of the mirrors as $Q_r=2 \pi N_{eff}/(1-|R|^2)$, where $|R|=\tanh(N \alpha |r_s|)$. Here $r_s$ is the reflection coefficient per electrode~\cite{Datta1986}, $\alpha$=0.2 is a scaling factor to account for the angle-dependent reflectivity, $N=200$ is the number of electrodes in the mirrors, and $N_{eff}\approx1/2|r_s|+7$ is the effective cavity length in units of the wavelength. We compute a quality factor associated with bulk loss as $Q_b=2 \pi N_{eff}/[C_b (h/\lambda)^2]$, where $C_b \approx 8.7$ is appropriate for LiNbO$_3$. Finally, we assume a frequency-independent quality factor associated with material loss (or a material-based dissipation rate that increases linearly with frequency) $Q_m \approx 1800$. We compute the total quality factor as  $Q=(Q_b^{-1}+Q_r^{-1}+Q_m^{-1})^{-1}$.
    
\section{Finite element simulation}
\begin{figure}
    \centering
    \includegraphics[width=0.75\textwidth]{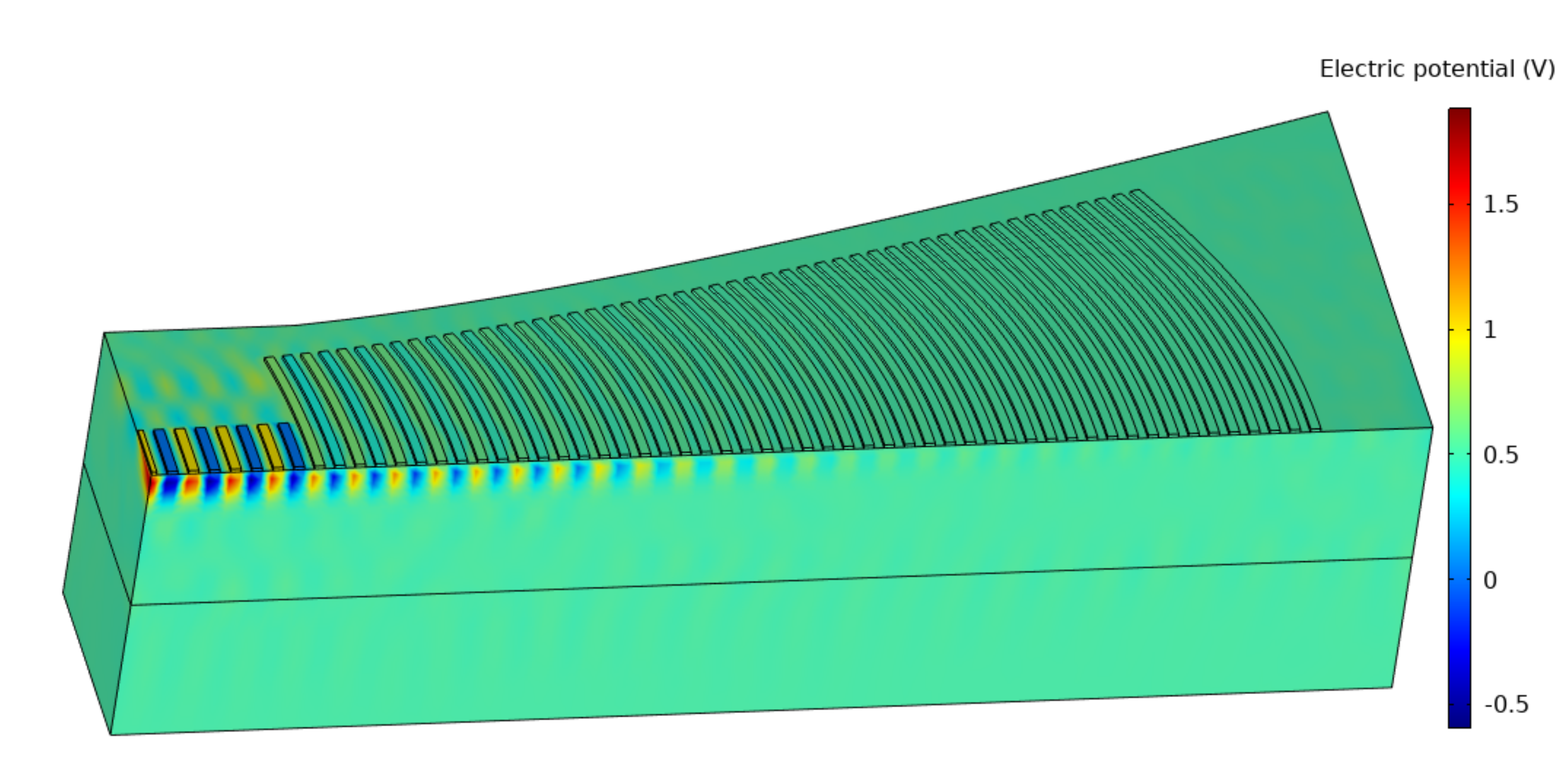}
    \caption{Example of a COMSOL model for simulating SAW resonators. The electric potential near resonance is shown.}
    \label{fig:comsol}
\end{figure}

We use COMSOL Multiphysics 5.6 to simulate our SAW resonators. To minimize the computational resources required, we simulate one-fourth of the full device and apply symmetric boundary conditions on the relevant faces of the model. The thickness of the substrate is 8$\lambda$, and the lower $4\lambda$ are a perfectly-matched layer. In all areas of the device except the perfectly-matched layer, the mesh size is less than or equal to $\lambda/4$. Figure~\ref{fig:comsol} shows a representative COMSOL model of our devices. The simulated devices differ from the actual devices in three ways. First, the electrode spacing in the simulated devices is not corrected for mass-loading effects. Second, the simulated devices include only 50 electrodes in each mirror (the actual devices contained 200). Finally, the simulated devices do not include dummy electrodes. Otherwise, the simulated devices are the same as the actual devices. 

\begin{figure}
        \centering
        \includegraphics[width=1.0\textwidth]{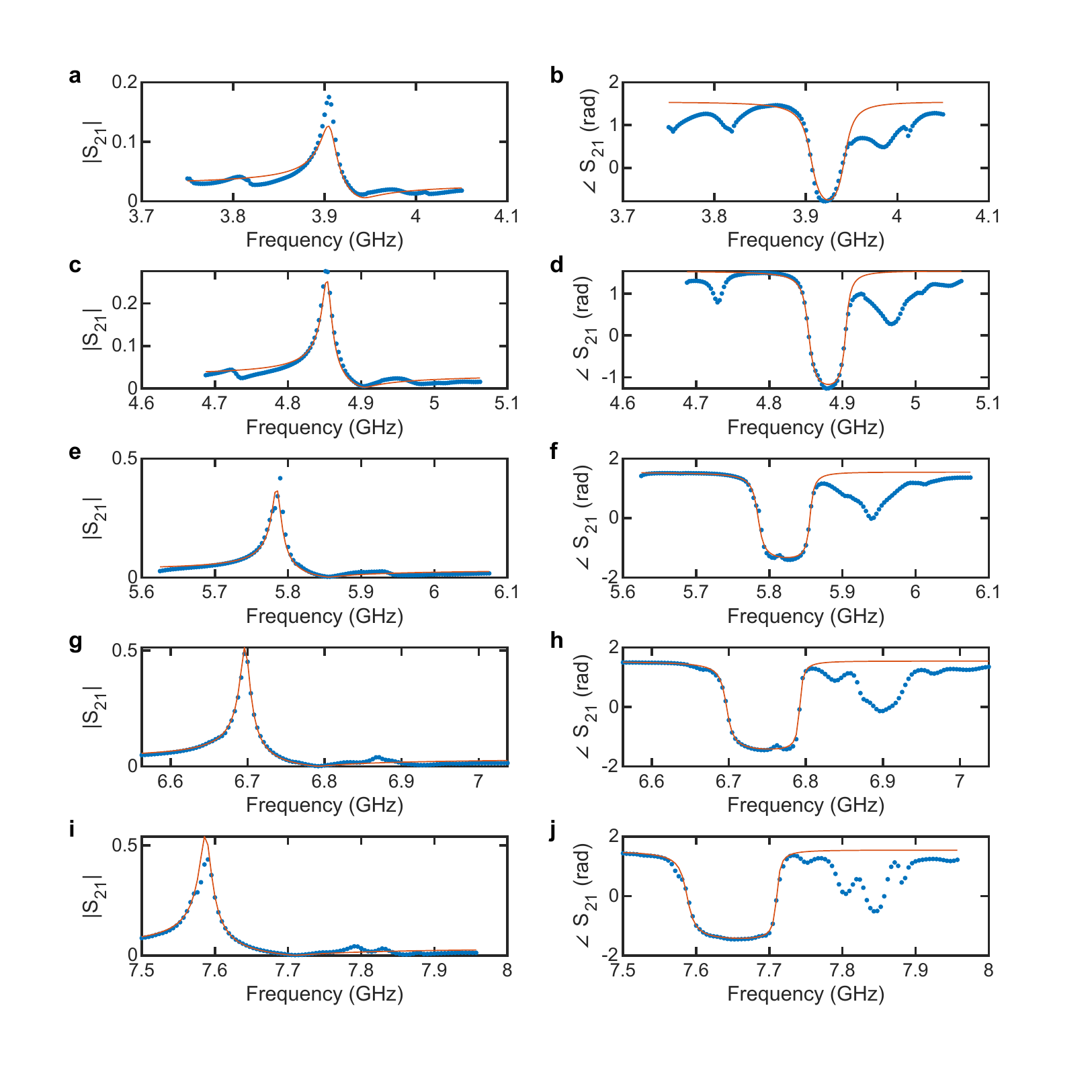}
        \caption{Data and fits from COMSOL simulations.
        (a)-(b) Simulated transmission and fits for a $\sim$4 GHz resonator. 
        (c)-(d) Simulated transmission and fits for a $\sim$5 GHz resonator. 
        (e)-(f) Simulated transmission and fits for a $\sim$6 GHz resonator. 
        (g)-(h) Simulated transmission and fits for a $\sim$7 GHz resonator. 
        (i)-(j) Simulated transmission and fits for a $\sim$8 GHz resonator. 
        In all panels, points are the data, and solid lines are fits. The fit results are summarized in Fig. 3(a) in the main text.}
        \label{fig:comsoldata}
    \end{figure}

The simulations were run on a cluster node with 56 cores and 512 GB RAM, and we extract the admittance $Y$ through the transducer as a function of frequency. From the simulated admittance values, we compute the impedance $Z=1/(4Y)$ and the transmission $S_{21}=2Z_0/(Z+2 Z_0)$, where $Z_0=50~\Omega$. The factor of 4 in the denominator is necessary because we simulate only 1/4 of the actual devices. We then fit the magnitude and phase of the transmission to a model with a parallel RLC circuit with a coupling capacitor to extract the circuit parameters. Figure~\ref{fig:comsoldata} shows all of the data and fits

\section{Flat resonators}
\begin{figure}
    \centering
    \includegraphics[width=1.0\textwidth]{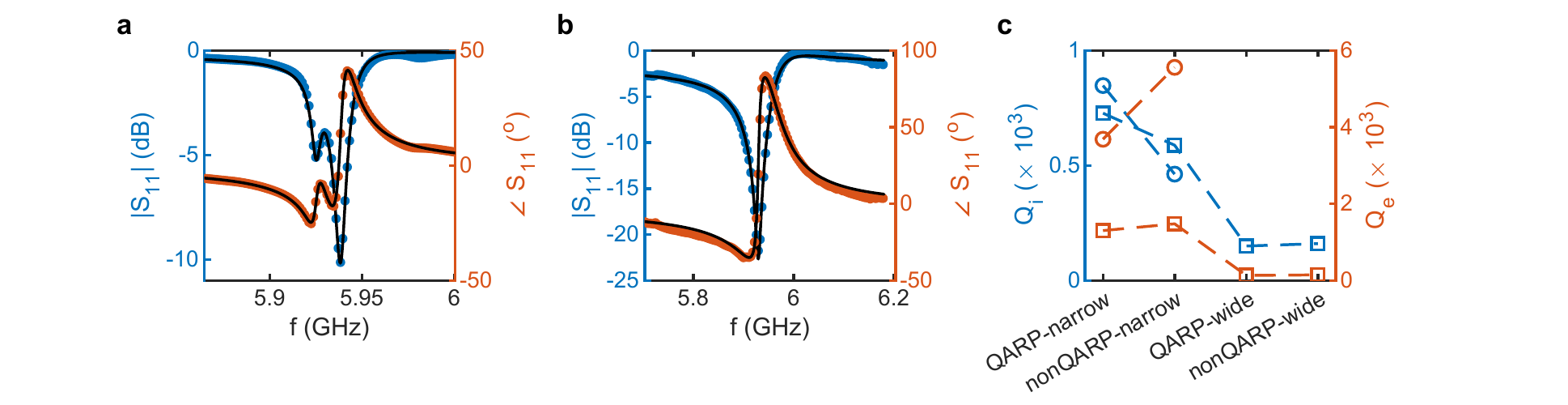}
    \caption{ Data and fits to the reflection mode model for $\sim$6-GHz flat resonators.
    (a) $S_{11}$ data and fits for a narrow flat resonator (width $4\lambda$) with QARP geometry. We observe double modes on all narrow flat mirrors we tested. 
    (b) $S_{11}$ data and fit for a wide flat resonator (width $42\lambda$) with QARP geometry.
    (c) The internal and external quality factors for flat resonators with different designs where squares are for the prominent mode and the circles are for the less prominent mode.}
    \label{flatmirror}
\end{figure}

We made $\sim$6-GHz flat resonators with different widths along the transverse direction, and with/without the QARP geometry. Typical reflection measurements from the flat resonators, fits to a reflection-mode model~\cite{mcrae2020materials}, and the results of the fits for different designs are shown in Fig.~\ref{flatmirror}. The ``narrow" and ``wide" flat resonators are 4$\lambda$ and 42$\lambda$ wide along the transverse direction. All narrow flat resonators we tested have multiple modes with internal quality factors lower than those of corresponding Gaussian resonators. Although one might expect a higher internal quality factor for wider resonators due to reduced diffraction losses, comparing the internal quality factors between narrow and wide resonators is difficult because of the presence of multiple apparent modes in the narrow resonators. The reduction in the external quality factor for the wide resonators occurs because of the larger transducer capacitance. 
\section{Cryogenic data}
\begin{figure}
    \centering
    \includegraphics[width=0.75\textwidth]{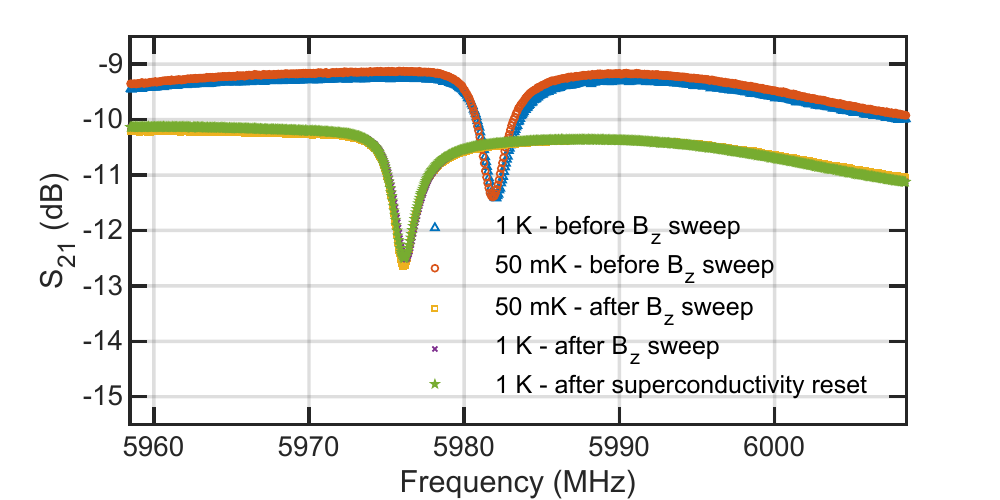}
    \caption{Data for $\sim$6 GHz resonator before and after the in-plane magnetic field ($B_z$) sweep. The resonance frequency and the background level irreversibly shifted downward at 1.3 T. The green and purple data are nearly identical.}
    \label{fig:Raw_Data_Comparison}
\end{figure}

While ramping the in-plane magnetic field, we observed an irreversible downward frequency jump ($\sim 3-6.5$ MHz) in all resonators at $\sim$1.3 T. The raw data for the $\sim$6 GHz resonator at different temperatures, and before and after the magnetic-field-sweep experiments, are shown in Fig.~\ref{fig:Raw_Data_Comparison}. The frequency of the resonators after the switch remained stable and did not return to the pre-switch values even after heating the substrate above the critical temperature of the Nb transmission line. We hypothesize that field-induced changes in either our cryogenic circulator or the lithium niobate substrate may explain these results. A detailed study and explanation of this observed permanent resonance frequency shift at certain magnetic fields is left for future research.

%SAWR-DOT ckt
\begin{figure}
    \centering
    \includegraphics{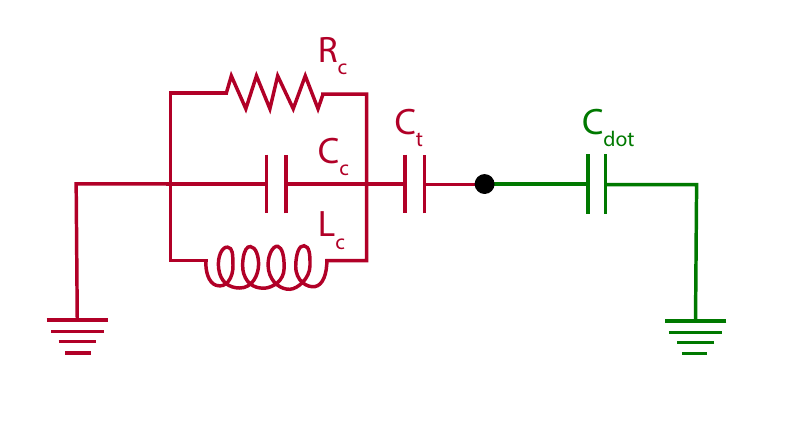}
    \caption{An ideal classical circuit diagram illustrating a SAWR coupled to a gate-defined quantum dot. The semiconductor quantum dot is modeled as a capacitor and is coupled to the resonator via the IDT capacitor.  Typical values for C$_{dot}$ are in the range of 10s of aF~\cite{Kouwenhoven2001}. The values of resonator circuit parameters are shown in Figs.~\ref{fig:fitParamsForFig2cd} and \ref{fig:pBVDparamsForQARP-nonQARP}. In this example of SAWR-quantum dot coupling, the SAWR voltage couples electrostatically to the spin or charge degrees of freedom in the quantum dot. The experimental implementation of this circuit can introduce parasitic capacitances which have detrimental effects on the coupling. As a general guideline, parasitic capacitances should be smaller than the IDT capacitance to minimize the effect on the coupling}
    \label{fig:sawr_dot_ckt}
\end{figure}

%\bibliography{SAWbib.bib}	